\documentclass[prl,aps,twocolumn,superscriptaddress,showpacs]{revtex4}
\usepackage{graphicx}

\begin{document}

\title{Diffusion-limited deposition of dipolar particles}

\author{F. de los Santos}
\affiliation{Departamento de Electromagnetismo y F\'\i sica de la Materia,
Universidad de Granada, Fuentenueva, 18071 Granada, Spain}

\author{J. M. Tavares}
\affiliation{Centro de F\'{\i}sica Te\'{o}rica e Computacional
da Universidade de Lisboa \\
Avenida Professor Gama Pinto 2, P-1649-003 Lisbon, Portugal}
\affiliation{Instituto Superior de Engenharia de Lisboa\\ Rua Conselheiro 
Em\'{\i}dio Navarro, 1, P-1949-014 Lisbon, Portugal}

\author{M. Tasinkevych}
\affiliation{Max Planck Institute for Metal Research,
Heisenbergstrasse 3, 70569 Stuttgart, Germany
}
\author{M. M. Telo da Gama}
\affiliation{Centro de F\'{\i}sica Te\'{o}rica e Computacional
da Universidade de Lisboa \\
Avenida Professor Gama Pinto 2, P-1649-003 Lisbon, Portugal}
\affiliation{Departamento de F\'{\i}sica, Faculdade de Ci\^encias da
Universidade de Lisboa, \\
R. Ernesto Vasconcelos, Lisbon, Portugal}

\date{today}

\begin{abstract}
Deposits of dipolar particles are
investigated by means of extensive Monte Carlo simulations.
We found that the effect of the interactions is described by an
initial, non-universal, scaling regime
characterized by orientationally ordered deposits.
In the dipolar regime, the order and geometry of the
clusters depend on the strength of the
interactions and the magnetic properties are tunable
by controlling the growth conditions.
At later stages, the growth is dominated by thermal effects and the
diffusion-limited universal regime obtains, at finite temperatures.
At low temperatures the crossover size increases exponentially as $T$
decreases and at $T=0$ only the dipolar regime is observed.
\end{abstract}


\maketitle

The growth of deposits by irreversible aggregation of particles is of great 
technological importance as well as of theoretical 
interest. A variety of mechanisms are involved in the growth 
processes but at late times scaling laws depending only on a few
parameters have been observed. A general assumption, that 
describes the patterns found in many experiments, is that the deposition 
process is dominated by thermal diffusion. 
A simple model for this type of growth is {\it diffusion-limited deposition} 
(DLD) \cite{meakin1}, characterized by the formation of branched, fractal 
structures similar to those found in electrodeposition, dielectric breakdown, 
etc (see \cite{reviews} and references therein). 
Under certain circumstances, however, interparticle interactions, favoring 
ordered structures that compete with the randomness of the diffusion process, 
are required to describe the observed growth patterns. A case in point is 
the diffusion-limited deposition of magnetic particles subject to dipolar 
interactions.

Dipolar interactions are essential in determining the
rich variety of structures exhibited by magnetic materials 
\cite{mm}, and their interplay with thermal diffusion may lead to
novel magnetic properties.
On the theoretical side, dipolar interactions provide a simple model to 
study the effects of anisotropic, long-ranged interactions on far from 
equilibrium aggregation processes.
The central question concerns the change in the fractal dimension of the
aggregates, $D$, as the dipolar interactions are switched on. 
Meakin et al. \cite{meakin2} considered the effect of
isotropic long-ranged, $1/r^\epsilon$, interactions in 
reaction-limited cluster-cluster aggregation (CCA) models. They found that 
$D$ is unchanged for short-ranged interactions, i.e. for $\epsilon \ge 
2D_o$ where $D_o$ is the fractal 
dimension of the non-interacting aggregates, while for longer-ranged 
interactions $D$ may change substantially. 
Accordingly, numerical results for diffusion-limitted aggregation (DLA),
performed for particles with Ising spins (short-range interactions),
revealed no changes in the fractal dimension of the aggregates with
increasing exchange interactions \cite{vandewalle}. Finally,
results for DLA \cite{rubis} of dipolar particles indicate that 
$D$ decreases as the
strength of the dipolar interactions increases, in line with
the results for CCA of dipolar particles \cite{mors} and with experimental
results for the aggregation of magnetic microspheres \cite{helgesen}, 
but in disagreement with preliminary results of ours \cite{nos1}.

In this article we report results of extensive Monte Carlo simulations 
that provide a general framework where the apparently contradictory 
results described above may be understood. We show that the
initial stage of two dimensional dipolar DLD growth is indeed described 
by a new nonuniversal scaling regime, characterized by clusters 
(trees) whose shape and fractal dimension are temperature-dependent. 
For large enough systems, however, the dipolar regime crosses-over to
the diffusion driven universal regime, where the effect of the dipolar 
interactions is dominated by thermal effects. It is also shown that 
the dipolar regime corresponds to orientationally ordered deposits and 
that the onset of the universal regime coincides with the disappearance 
of the orientational order. At $T=0$ only the dipolar regime is observed. 

In the new dipolar regime, the orientational order as well as the shape 
and fractal dimension of the clusters depend on the strength of the
interactions. Thus, the magnetic properties of dipolar deposits may be 
tuned by controlling the growth conditions, such as temperature. 
Finally, we found that the fractal dimension of the entire deposit is 
always given by the universal (diffusion driven) value implying 
that in the dipolar regime, the trees have a fractal dimension $D_t$ that 
differs from that of the entire deposit.

\begin{figure}
\begin{center}
\includegraphics[width=7cm]{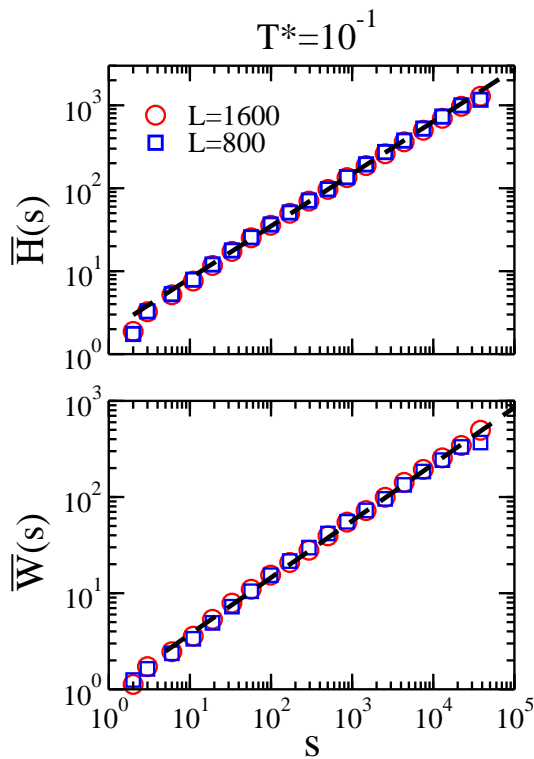}
\end{center}
\vskip-0.5cm 
\caption{
(Color online) Average height ${\bar H}(s)$ and width 
${\bar W}(s)$
as a function of the size of the tree, $s$, at $T^*=10^{-1}$ and various  
system sizes. 
The lines are power laws (\ref{scahw}) with the exponents 
listed in table I. Dashed line: universal regime. 
}
\end{figure}

\begin{figure}
\begin{center}
\includegraphics[width=7cm]{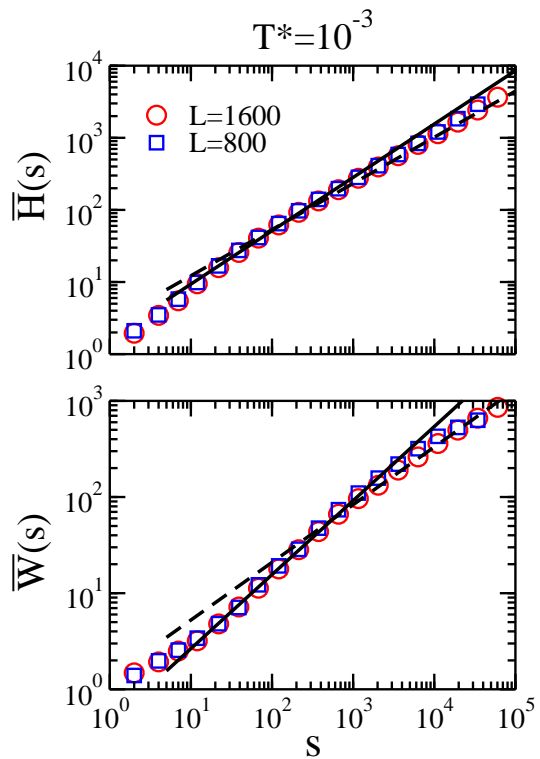}
\end{center}
\vskip-0.5cm 
\caption{
(Color online)
Average height ${\bar H}(s)$ and width 
${\bar W}(s)$
as a function of the size of the tree, $s$, at $T^*=10^{-3}$
and various system sizes. 
The lines are power laws (\ref{scahw}) with the exponents 
listed in table I. Full line: 
dipolar regime. Dashed line: universal regime. 
}
\end{figure}

\begin{figure}
\begin{center}
\includegraphics[width=7cm]{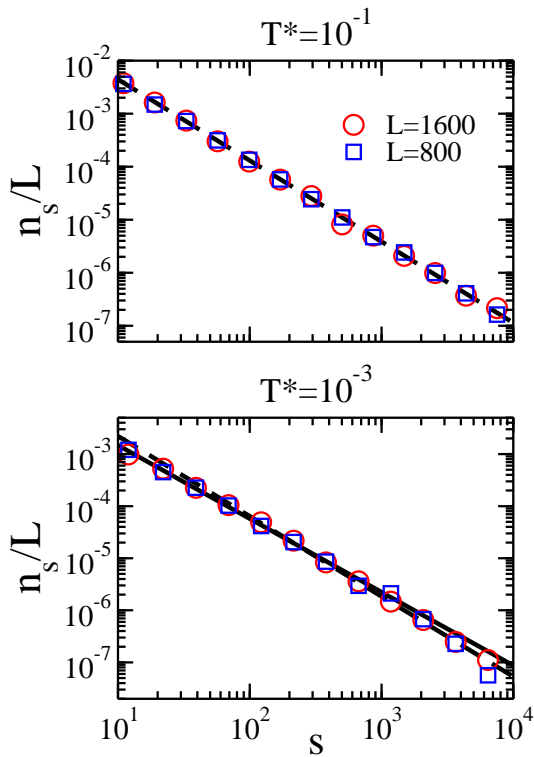}
\end{center}
\caption{
(Color online)
Average number of trees of size $s$ divided by $L$ (${\bar n}_s/L$) as 
a function of $s$, at various temperatures and system sizes. 
Lines as in figure 2.}
\end{figure}

We consider a two-dimensional box of side $L$ and height $H$, on a
square grid of side $a=1$. Periodic  
boundary conditions are applied parallel to the substrate, of size $L$, and 
particles are dropped from a finite height 
above it. The particles carry a dipole moment of strength $\mu$ and interact 
through the pair potential
\begin{equation}
\phi_{1,2}= -{\mu^2 \over r_{12}^3} 
\Big[3
(\hat {\mbox{\boldmath$\mu$}}_1 \cdot {\hat {\bf r}}_{12})
(\hat {\mbox{\boldmath$\mu$}}_2 \cdot {\hat {\bf r}}_{12})
-\hat {\mbox{\boldmath$\mu$}}_1 \cdot \hat {\mbox{\boldmath$\mu$}}_2
\Big],
\end{equation} 
where $r_{12}$ is the distance between particles 1 and 2, 
$\hat {\bf r}_{12}$ is the two-dimensional unit vector along 
the interparticle axis, and $\hat {\mbox{\boldmath$\mu$}}_1$ and 
$\hat {\mbox{\boldmath$\mu$}}_2$ are the three-dimensional unit vectors in
the direction of the dipole moments of particles 1 and 2 respectively. 
A particle is released at a height $H_{in}$
with a dipole moment oriented at random. 
The particle undergoes a random walk 
through a series of jumps to nearest-neighbor sites, while interacting
with the particles in the deposit. 
At each step a new position and a new random 3-dimensional dipole orientation 
are accepted according to a simple 
Metropolis rule based on the difference between the dipolar energies of the 
two configurations and defining the effective temperature
$T^*=k_BTa^3/\mu^2$.  
The long range of the dipolar interactions is taken into account by an 
Ewald summation for the slab geometry of the system \cite{nos1}.
In the limit $T^*\to \infty$ all displacements are accepted
and the model reduces to DLD.
A particle will eventually ({\it i}) contact the deposit or the substrate
sticking to it irreversibly as its dipole relaxes along the local field
\cite{relaxation}; or ({\it ii}) reach a height greater than $H_{out}$,
when it is removed and a new particle released. See \cite{nos1,nos2} for
the details of the simulation. 

Simulations were carried out at 4 temperatures,
$T^*=10^{-1}, 10^{-2}, 10^{-3}, 10^{-4}$,
and 4 system-sizes, $L=200, 400 , 800, 1600$ with
20, 30, 50, and 100 thousand particles per deposit,
respectively. The deposits are similar to those of DLD: they
consist of many small trees competing to grow \cite{nos1}. As the number
of particles in the deposit increases, fewer and fewer trees
keep on growing due to shadowing until only a single tree
survives.

We start by investigating the dependence of the height, $H$, and the width, 
$W$, of a tree with its size $s$ (number of particles), as well as the 
distribution of trees $n_s$, i.e. the average number of trees of a given size. 
In DLD the trees scale as \cite{reviews,meakinandracz,meakin3},
\begin{equation}
H \sim s^{\nu_\parallel}, \quad W \sim s^{\nu_\perp}, \quad n_s 
\sim s^{-\tau}.
\label{scahw}
\end{equation}
for sufficiently large values of $s$. Owing to the finite size of the 
simulation box, however,
only a single tree survives when the number of deposited particles 
is large enough and, thus
at large $s$ the width of the tree saturates ($W \approx L$), then the 
height grows linearly with its size, ($H \propto s$), and $n_s$ exhibits a
discontinuity.

The difference between $\nu_{\parallel}$ and $\nu_{\perp}$ measures 
the anisotropy of the trees. If $\nu_{\parallel}=\nu_{\perp}$ the trees 
are isotropic and their fractal dimension is 
$D_t=1/\nu_\parallel$, whereas if anisotropy is present, $D_t$ 
becomes \cite{reviews}
\begin{equation}
D_t=1+(1-\nu_\parallel)/\nu_\bot.
\label{eq:hypersc2}
\end{equation}
We note that the assumption that the deposit has the same fractal
dimension as the trees, $D=D_t$, 
which holds in DLD, is not warranted in general.  
$D$ was estimated through
the average particle density at height $h$, $\rho(h)$, which was 
found to scale as in
DLD: 
at early times the deposit builds up until it reaches
a height $h_i$. Then, we found a scaling
regime where the density decreases as a power law 
of the height, $\rho(h) \sim h^{-\alpha}$, with $D=2-\alpha$ \cite{reviews}.
The density saturates when the lateral correlation 
length $\xi_\parallel$ reaches the size of the system,  
at $h_s \sim L^\gamma$. 
Simple arguments show \cite{reviews} that the 
exponents $\nu_{\parallel}$, $\tau$ and $\alpha$ are related 
through  
\begin{equation}
\label{eq:hypersc1}
\alpha= 1- (2-\tau)/\nu_{\parallel}.
\end{equation}
We have verified (\ref{scahw}) by 
calculating the average maximal height and width 
of the largest tree,
${\bar H}(s)= \langle h(s) \rangle$,
${\bar W}(s)= \langle w(s) \rangle$,
as a function of the tree size, $s$,
and ${\bar n}_s  = \langle n_s \rangle$.
$h(s)$ and $w(s)$ are the maximal height and width of the largest tree 
of each deposit, and $\langle \ldots \rangle$ is an
average over all the deposits. 

Figures 1 and 2 show the results for ${\bar H}(s)$
and ${\bar W}(s)$ obtained at $T^*=10^{-1}$ and $T^*=10^{-3}$ for different 
box sizes. 
The points are histograms on a logarithmic scale. 
As expected, the results at $T^*=10^{-1}$ correspond to those of DLD. 
Scaling was found for a wide range of $s$ ($s>10$), after a crossover
from the early stage regime ($s<10$).  
The crossover to the linear regime (only-one-tree growth) was observed
only for the smallest system, $L=200$, for $s>5000$ (not shown). 
$nu_\parallel$ and $\nu_\bot$ calculated for each L were found independent of system
size; a representative value was obtained by averaging over all $L$ with the
uncertainty estimated as the largest deviation from the mean.
Power law regressions for different ranges of $s$, 
yield $\nu_{\parallel}=0.64(1)$. 
The crossover to saturation of the width 
was observed for sizes up to $L=800$. 
Fitting only points that exhibit clear data
collapse yields $\nu_{\perp}=0.60(2)$. These
exponents agree with previous results for DLD \cite{meakinandracz}, and
show that at high 
temperatures, the universal behavior of the geometrical properties of the 
trees is unaffected by the dipolar interactions.  

At $T^*=10^{-3}$, however, the existence of two scaling 
regimes is apparent, in particular for the largest system, $L=1600$, 
where the width saturation and the linear regimes are not observed.  
For $s$ less than the crossover size $s^* \approx 500$ we found
$\nu_\parallel = 0.74(1)$ and $\nu_\bot = 0.78(2)$. These exponents
differ from those of DLD and characterize a new growth regime that 
we call the {\em dipolar} regime.
For $s> s^*$, we obtained, $\nu_\parallel = 0.64(1)$ and 
$\nu_\bot = 0.60(2)$, in line 
with the results for DLD. 
Similar behavior was observed at the other two 
temperatures, that is, 
a dipolar regime with temperature-dependent exponents followed by 
a second scaling regime with DLD exponents. 

In table I we list the exponents of the 
dipolar and universal regimes, obtained at various temperatures. 
These results indicate that, if (\ref{scahw}) is assumed,  
the effect of the dipolar interactions may be described by the appearance
of a dipolar regime characterized by non-universal exponents 
$\nu_{\parallel}$ and $\nu_\perp$ that increase with
decreasing temperature. The crossover to the universal regime 
occurs at tree sizes that increase as the temperature decreases. 
As a consequence, the universal regime is difficult to observe 
at very low temperatures. 
However, at finite $T^*$, the universal regime may be reached if large
enough
deposits are grown. Indeed, we found no dependence of the crossover 
between the dipolar and the universal regimes on $L$, and thus at 
any temperature there is an $L$ above which this crossover may be
observed.  
This implies that the geometrical properties of the trees may be tuned by
controlling the
dipolar interactions and the system size. In particular, it is possible to
deposit trees with a given anisotropy by growing trees (at a fixed
temperature) with a fractal dimension that is determined by the tree size. 
Alternatively, at fixed tree size, one may control the anisotropy by
changing the temperature.

\begin{table*}
\caption{Characteristic exponents of the dipolar DLD model. 
$\nu_{\parallel}$, $\nu_{\perp}$ and $\tau$ obtained from the simulation; 
other exponents obtained using the equations indicated in 
brackets.} 
\begin{ruledtabular}
\begin{tabular}{ccccccccc}
$\,T^*\,$&regime& $\nu_{\parallel}$& $\nu_{\perp}$&$\tau$&$\alpha$&
$\tau$ (\ref{eq:hypersc1})&
$D_{t}$ (\ref{eq:hypersc2})& $D_t\equiv1/\nu_{\parallel}$\\
\hline
$\infty$(DLD)&universal&0.630(2)&0.580(4)&1.56(2)&0.288(2)&1.551(3)&
1.64(1)&1.59(1)\\
$10^{-1}$& universal&0.64(1)&0.60(1)&1.54(2)&0.29(1)&1.55(1)&1.60(3)&
1.56(2)\\
$10^{-2}$&dipolar&0.70(3)&0.69(3)&1.46(3)&0.27(1)&1.49(3)&1.43(8)&
1.43(6)\\	
$10^{-2}$&universal&0.63(2)&0.60(4)&1.53(4)&0.30(1)&1.56(2)&1.63(7)&
1.59(5)\\
$10^{-3}$&dipolar&0.75(2)&0.77(2)&1.40(2)&0.24(1)&1.44(2)&1.32(5)&
1.33(4)\\
$10^{-3}$&universal&0.64(1)&0.60(2)&1.56(4)&0.28(1)&1.55(1)&1.60(3)&
1.56(2)\\
$10^{-4}$&dipolar&0.83(3)&0.83(3)&1.36(3)&0.25(1)&1.38(3)&1.20(7)&
1.20(4)\\
\end{tabular}
\end{ruledtabular}
\end{table*}

In figure 3, we plot the results for the tree distribution 
at $T^*=10^{-1}$ and $T^*=10^{-3}$.
At $T^*=10^{-1}$, the results are those of DLD. Scaling behavior 
was observed for a range of $s$ between
$10$ and the maximum tree size (which depends on $L$), with 
an exponent $\tau = 1.54(2)$ in line with results for DLD 
\cite{meakinandracz}. 
At $T^*=10^{-3}$, assuming the crossover to occur 
at $s^*\approx 500$, we estimated $\tau$ in 
the ranges $10<s<s^*$ (dipolar regime) and $s^*<s<5000$ (universal regime). 
We found 1.40(2) and 1.56(4), respectively, for both systems. Thus,
as for the tree height and width, we found a dipolar 
regime for the tree distribution, characterized by a temperature-dependent 
exponent.
Again, the crossover to the universal regime occurs for sizes 
that increase as the temperature decreases. 
This crossover was not observed at $T^*=10^{-4}$: due to the 
limited number of deposits and decreasing density,  
the number of trees larger than $s \approx 500-1000$ was too small to
be analyzed. Thus, at this temperature only the dipolar regime was
observed, with $\tau = 1.36(3)$).

We have used (\ref{eq:hypersc2}) and (\ref{eq:hypersc1}) to verify the 
consistency of the exponents and to estimate $D_t$. In addition, we have 
checked the validity of   
(\ref{eq:hypersc1}) in the dipolar regime and compared $D_t$ in both 
regimes. 
$\alpha$ was estimated in both regimes by power law regressions of
$\rho(h)$ 
in the regions suggested by the plots of ${\bar H}_l$ 
(using data from $L=1600$). 
We found weak crossovers at the temperatures $T^*=10^{-2}$ and $10^{-3}$, 
similarly to what was observed for ${\bar n}_s$.  
The results listed in Table I indicate a remarkable
consistency between the 
values of $\tau$ obtained from simulation 
and using (\ref{eq:hypersc1}).   
It is also clear that the fractal dimension of the trees in the 
dipolar regime decreases with decreasing temperature, 
in line with previous studies of DLA dipolar aggregates
\cite{rubis,mors}.  

\begin{figure}
\begin{center}
\includegraphics[width=6cm,angle=270]{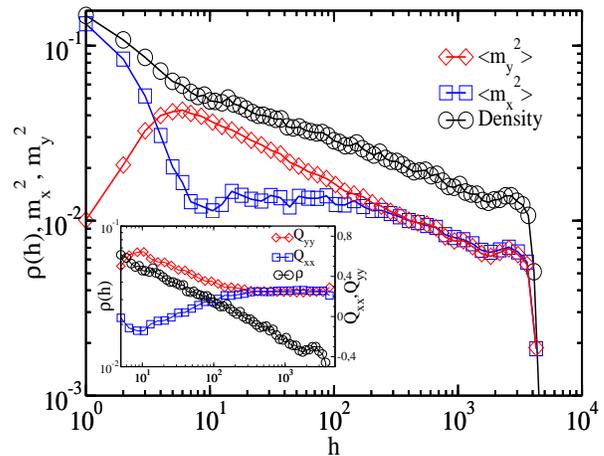}
\end{center} 
\vskip-0.5cm
\caption{
(Color online)
Average particle density and mean-square magnetization densities
at height $h$. Inset: diagonal elements, $Q_{xx}$ and $Q_{yy}$, 
of the ordering matrix $\bf Q$.
}
\end{figure}

The connection between the orientational order of the dipoles 
and the geometrical properties of the deposits was investigated by 
comparing the mean-square magnetization density in the 
$x-$ and $y-$ directions at height $h$, $\langle m_{x,y}^2(h) \rangle$,
with the mean particle density at the same height, $\rho(h)$.
These results are shown in figure 4.
It is apparent that the density 
saturates at the same height as the mean-square magnetizations, and that the 
onset of the 
scaling behavior of $\rho$ coincides with that of $\langle m_y^2 \rangle$.
Notice also the height $h^\dagger \approx 300$, where $\langle m_x^2
\rangle
= \langle m_y^2 \rangle$ and the orientational order vanishes. 
That $h^\dagger$ signals the disappearance of orientational order is seen
most clearly in the 
inset of figure 3. The latter shows that the diagonal elements,
$Q_{xx}(h)$ and $Q_{yy}(h)$, 
of the ordering matrix ${\bf Q}$ become identical at 
$h=h^\dagger$, beyond which $Q_{xx} = Q_{yy} = 1/4$ 
\cite{nos2}. 
We also found that $h^\dagger$ does not depend on $L$,
but increases with decreasing $T^*$. At low $T^*$, the increase is
exponential, $h^\dagger \simeq \exp(1/T^*)$.
This functional dependence was obtained from a direct estimation
of the heights at which the orientational order of the deposits
vanishes. 
Depending on the temperature and system size, $h^\dagger$ is larger or
smaller than $h_s$ and this is related to the existence of the dipolar and 
universal regimes: when $h^\dagger < h_s$ both the universal and the dipolar 
regimes occur, but only the latter is found when $h^\dagger >h_s$. 
We recalculated the
scaling exponents assuming that the crossover between the
dipolar and universal regimes occurs at $h=h^\dagger$ (equivalently, 
at $s^{\dagger}$ estimated using ${\bar H}(s)$) and found that they are
identical with those of table I. Therefore, the dipolar 
regime is characterized by deposits with orientational order.
As a consequence, and by contrast to what happens in DLD, the fractal
dimension of the deposits depends both on the fractal dimension of the trees,
$D_t$, and on the distribution of the trees, $\tau$. 
In the light of this, the decreasing $D_t(T^*)$ reported in 
\cite{mors,helgesen}, is to be expected, since it was determined from  
the radius of gyration versus the (small) number of particles of each 
cluster.
Furthermore, the results of \cite{rubis}, obtained for a single DLA
cluster, may be interpreted as the crossover from the
temperature-dependent fractal dimension at short
length scales, to $D\simeq 1.7$ at long length scales \cite{rubis}.

We have also estimated the average interaction of a dipolar 
particle with similar particles in ordered and randomly oriented deposits 
numerically, and confirmed that for ordered deposits the interaction 
decays more slowly than 
$2D_o(\equiv2 \times D_t(T=\infty)\approx 3.28)$
while for random ones it decays faster, 
in line with the results of \cite{meakin2} for isotropic systems \cite{nos3}.
Finally, we have checked that the global fractal dimension
of the deposits is unaffected (or very weakly affected) 
by the dipolar interactions.
This was found by analyzing the decay of the mean density with $h$,
the scaling of $h_s$ and of the density at 
saturation with $L$,
the two-point density-density correlation function, the initial 
divergence of the interface width, and the mean height of the upper surface.
In every case no significant deviation from DLD was found \cite{nos3}.

\begin{acknowledgments}
Funding from the Portuguese FCT, contract SFRH/BPD/5654/2001, 
and from the Spanish MCyT, project BMF2001-2841, is gratefully acknowledged.
\end{acknowledgments}


\begin{thebibliography}{99}

\bibitem{meakin1}
P. Meakin, Phys. Rev. A {\bf 27}, 2616 (1983).

\bibitem{reviews}
P. Meakin, {\it Fractals, scaling and growth far from equilibrium
}, Cambridge Universtiy Press, Cambridge (1998);
T. Vicsek, {\it Fractal growth phenomena}, World Scientific (1989).

\bibitem{mm}
K. De'Bell, A.B. MacIsaac, and J.P. Whitehead,
Rev. Mod. Phys. {\bf 72}, 225 (2000). 

\bibitem{meakin2}
P. Meakin and M. Muthukumar,
J. Chem. Phys. {\bf 91}, 3212 (1989).

\bibitem{vandewalle}
N. Vandewalle and M. Ausloos, Phys. Rev E {\bf 51}, 597 (1995). 

\bibitem{rubis}
R. Pastor-Satorras and J.M. Rub\'{\i},
Phys. Rev. E {\bf 51}, 5994 (1995);
Prog. Colloid. Polym. Sci. {\bf 110}, 29 (1998);
J. Magn. Magn. Mater. {\bf 221}, 124 (2000).

\bibitem{mors}
P.M. Mors, R. Botet, and R. Jullien,
J. Phys. A {\bf 20}, L975 (1987).

\bibitem{helgesen}
G. Helgesen, A.T. Skjeltorp, P.M. Mors, R. Botet, and R. Jullien,
Phys. Rev. Lett. {\bf 61}, 1736 (1988).

\bibitem{nos1}
F. de los Santos, M. Tasinkevych, J.M. Tavares, and P.I.C. Teixeira,
J. Phys.: Condens. Matter {\bf 15}, S1291 (2003).

\bibitem{relaxation}
The relaxation of the dipole does not appear to be an essential ingredient
of the algorithm, since the same $D$ has been found in \cite{mors}
with and without dipole moment relaxation. 

\bibitem{nos2} 
J.M. Tavares, M. Tasinkevych, F. de los Santos, and M.M. Telo da Gama,
Mol. Phys. {\bf 101} 1659 (2003).

\bibitem{meakinandracz}
P. Meakin, Phys. Rev. B {\bf 30}, 4207 (1984);
P. Meakin, J. Kert\'esz, and T. Vicsek, 
J. Phys. A {\bf 21}, 1271 (1988);
Z. R\'acz and T. Vicsek,
Phys. Rev. Lett. {\bf 51}, 2382 (1983).


\bibitem{meakin3}
P. Meakin and F. Family, Phys. Rev. A {\bf 34}, 2558 (1986).

\bibitem{nos3}
J.M. Tavares et al., unpublished.
\end{thebibliography}
\end{document}